\documentclass[11pt]{iopart}

\pdfoutput=1
\usepackage{natbib}

\usepackage{newblock}

\usepackage{graphicx}
\usepackage{siunitx}

\begin{document}

\title[3D reconstruction of carbon ion beams from secondary particle tracks]{Three dimensional reconstruction of therapeutic carbon ion beams in phantoms using single secondary ion tracks} 

\author{Anna Merle Reinhart$^1$, Claudia Katharina Spindeldreier$^1$, Jan Jakubek$^2$ and Maria Martisikova$^{1,3}$}

\address{$^1$ Medical Physics in Radiation Oncology, German Cancer Research Center (DKFZ), Im Neuenheimer Feld 280, 69120 Heidelberg, Germany}

\address{$^2$ Institute of Experimental and Applied Physics, Czech Technical University in Prague, Horska 3a/22, 12800 Prague 2, Czech Republic}

\address{$^3$ Department of Radiation Oncology, Heidelberg University Hospital, Im Neuenheimer Feld 400, 69120 Heidelberg, Germany}
\ead{m.martisikova@dkfz.de}

\date{\today}

\begin{abstract} 
Carbon ion beam radiotherapy enables a very localised dose deposition. However, already small changes in the patient geometry or positioning errors can significantly distort the dose distribution. A live monitoring system of the beam delivery within the patient is therefore highly desirable and could improve patient treatment. We present a novel three-dimensional imaging method of the beam in the irradiated object, exploiting the measured tracks of single secondary ions emerging under irradiation.\\
The secondary particle tracks are detected with a TimePix stack, a set of parallel pixelated semiconductor detectors. We developed a three-dimensional reconstruction algorithm based on maximum likelihood expectation maximization. We demonstrate the applicability of the new method in an irradiation of a cylindrical PMMA phantom of human head size with a carbon ion pencil beam of \SI{226}{\mega\electronvolt\per\atomicmassunit}. The beam image in the phantom is reconstructed from a set of 9 discrete detector positions between \SI{-80}{\degree} and \SI{50}{\degree} from the beam axis. Furthermore, we demonstrate the potential to visualize inhomogeneities by irradiating a PMMA phantom with an air gap as well as bone and adipose tissue surrogate inserts.\\
We successfully reconstructed a three-dimensional image of the treatment beam in the phantom from single secondary ion tracks. The beam image corresponds well to the distribution expected from the beam direction and energy. In addition, cylindrical inhomogeneities with a diameter of \SI{2.85}{\cm} and density differences down to \SI{0.3}{\g\per\cm\cubed} to the surrounding material are clearly visualized.\\
This novel three-dimensional method to image a therapeutic carbon ion beam in the irradiated object does not interfere with the treatment and requires knowledge only of single secondary ion tracks. Even with detectors with only a small angular coverage, the three-dimensional reconstruction of the fragmentation points presented in this work was found to be feasible.
\end{abstract}

\maketitle 

\section{Introduction}
Compared to conventional irradiations with photons, heavy-ion radiotherapy offers a highly conformal dose deposition to the target volume. This is due to the fact that ions deposit most energy in a very sharp, well-defined peak at the end of the ion range, the so-called Bragg peak. With an active beam delivery that scans thin ion pencil beams over the tumour volume, the dose deposition can be precisely tailored to the target and critical organs in the vicinity are spared. However, this strength of ion beam radiotherapy is also a challenge. Even small changes in the patient geometry such as tumour swelling or shrinking, organ motion, weight gain or loss, or small positioning errors can significantly distort the delivered dose distribution \citep{Enghardt2004}. The precise dose localization thus requires a highly accurate beam delivery and quality assurance. While the primary beam parameters such as its direction and the particle fluence can be verified online by an external beam monitoring system, the final dose distribution within the patient is much harder to quantify. A live, non-invasive monitoring system for the beam delivery in the patient is therefore highly desirable.

In this study, we concentrate on carbon ion radiotherapy. Besides dose deposition through coulomb interactions with electrons, the ions also interact with the nuclei, and target and projectile fragments are created. While target fragments predominantly stay at rest, projectile fragments have a velocity similar to the primary ion. Thus, most projectile fragments have a longer range in tissue than the primary particles due to their lower mass and charge. This causes a dose tail behind the Bragg peak, which is in the first place an unwanted effect. On the other hand, these secondary charged particles partly emerge from the patient. In addition, photons are produced either in prompt processes or in the de-excitation of radioactive fragments. If any of these by-products of the irradiation can be detected and analysed, they could offer additional information about the primary beam in the patient for free.

All main beam delivery monitoring techniques under development are based on the detection of secondary particles. The only treatment verification method currently clinically available is based on positron emission tomography (PET) \citep{Enghardt2004,Parodi2008}. Some of the created nuclear fragments are $\beta^+$ emitters and can thus be used for PET imaging. However, this technique suffers from several drawbacks. If the patient is moved from the treatment position to the PET scanner, physiological processes during the transition time lead to a washout of the $\beta^+$-emitting nuclei. In addition, the fast signal decay results in long measurement times to achieve reasonable counting statistics. In-beam PET is not affected by these processes, however it suffers from a high background and its installation is technically challenging and cost-intensive. \citep{Shakirin2011}.

To overcome these drawbacks, beam delivery verification based on the detection of prompt gammas \citep{Min2006,Testa2008,Testa2009,Testa2010,Smeets2012} and prompt secondary charged particles \citep{Amaldi2010,Henriquet2012,Agodi2012,Piersanti2014,Gwosch2013} has been proposed and is currently studied in several groups. These techniques potentially permit real-time monitoring of the dose deposition within the patient since the relevant physical processes take place on short time scales in the order of \SI{e-16}{\s} \citep{Kraft2001}. In contrast to PET monitoring, dedicated clinical detection systems for these kinds of radiation are not available yet.

Previous studies \citep{Agodi2012,Piersanti2014,Gwosch2013} have analysed two-dimensional projections of the beam image in the patient, thus losing information from the three-dimensional problem and increasing uncertainties. \cite{Gwosch2013}, for example, considered the intersection points of the measured particle trajectories with the beam plane. This requires an a-priori assumption of the position of the beam plane, and the neglected finite beam width smears out the distribution. A three-dimensional reconstruction of the fragmentation points does not suffer from these drawbacks. Two methods to gain 3D images were studied in \citep{Henriquet2012}, both based on the idea to reconstruct single fragmentation vertexes. While the vertex reconstruction based on two fragments originating from a single fragmentation event drops the detection statistics rapidly (by about one order of magnitude), finding an intersection point of a primary and secondary ion requires a primary ion detector in the beam which deteriorates the treatment beam quality. Therefore, another method for 3D beam imaging in the patient with high statistics and not interfering with the treatment is desired.

Our research group focusses on carbon ion beam monitoring by secondary charged particle tracking with the pixelated semiconductor detector TimePix \citep{Llopart2007}. The technology of particle tracking originally developed in high energy physics is applied to medical physics, where such an approach is not yet common. The TimePix detector can successfully register single ion tracks in carbon ion radiotherapy \citep{Jakubek2011b,Martisikova2011}. An initial study in a homogeneous polymethyl methacrylate (PMMA) phantom \citep{Gwosch2013} showed that monitoring of the beam width and position with sub-millimeter precision, as well as the detection of changes in the beam range down to \SI{1.3}{\mm} are possible.  

The work presented in this article aims to improve the above-mentioned 2D method. We developed a 3D beam image reconstruction based on single measured ion tracks and maximum likelihood expectation maximization (MLEM). This represents a new imaging modality of the beam in the irradiated object. Firstly, we present a proof-of-principle of this reconstruction in an irradiation of a homogeneous head-sized PMMA phantom with a therapeutic carbon ion beam. To demonstrate the capabilities of the novel method, we subsequently irradiated a phantom with inhomogeneities and investigated if the changes in the geometry are visible in the reconstructed images.


\section{Materials and methods}
\label{sec:Methods}

\subsection{Tracking of secondary ions with the TimePix detector}
\label{sec:Timepix}

The image reconstruction requires directional information of the individual secondary particle tracks, which we register with the TimePix detector \citep{Llopart2007}. The TimePix detector is a hybrid semiconductor pixel detector developed within the Medipix collaboration at CERN \footnote{http://medipix.web.cern.ch/medipix/}. It can detect individual ionizing particles with a very high efficiency \citep{Soukup2011}. The detector consists of a \SI{14 x 14}{\mm} semiconductor detector chip, bump-bonded to a pixelated readout chip with $256 \times 256$ square pixels of \SI{55 x 55}{\micro\meter} pitch \citep{Llopart2007}. In the measurements presented in this article, we used a sensitive layer of \SI{300}{\micro\meter} silicon.  Different operation modes allow for the measurement of either the particle arrival time, the energy deposition or the number of hits in a pixel \citep{Llopart2007}. In this work, we measured the particle arrival time. The signal is read out in so-called frames with a certain acquisition time that can be chosen by the user. A global shutter signal delimits the frames. The detector is controlled with the USB-based interface FITPix \citep{Kraus2011} and the software package Pixelman \citep{Turecek2011} is used for data acquisition.

Due to the charge-sharing effect \citep{Jakubek2009}, multiple adjacent pixels register a non-zero signal for a single hit, forming a so-called cluster \citep{Gwosch2013}. To guarantee a high quality of the data, we set constraints on the cluster parameters to exclude unwanted events. One-pixel events are attributed to photons or electrons and hence excluded from further analysis. To exclude malfunctioning pixels or particles arriving just before the start of the frame, clusters with a time stamp of the beginning of the frame are not analysed. 

In order to obtain information on the secondary particle trajectories, multiple detectors are combined to form a three-dimensional voxel detector \citep{Soukup2011}. Coincident events in the different detector layers can be matched. For all measurements in this study, the coincidence time limit was set to \SI{100}{\nano\second} (1 ADC count).

\subsection{Experiments}
\label{subsec:Experiments}

\begin{figure}
\centering
\includegraphics[width=.7\textwidth]{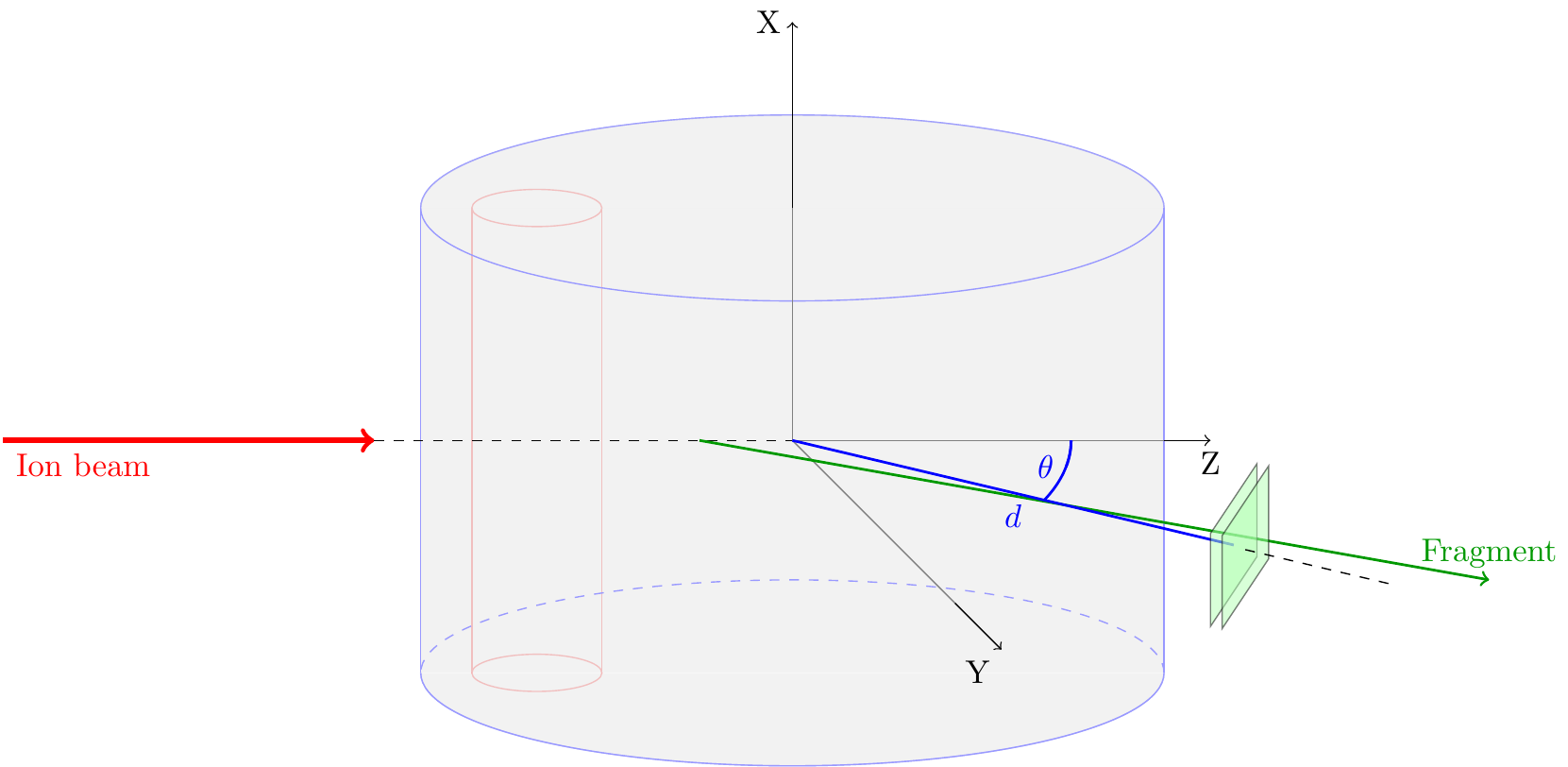}
\caption{Schematic illustration of the experimental setup. The incident ion beam (red) is directed along the z axis. The two parallel TimePix detectors are placed at distance $d$ and angle $\theta$ from the center of the phantom. Two different phantoms were employed, a cylindrical homogeneous PMMA phantom and a PMMA cylinder of the same outer dimensions as the homogeneous phantom, but with a cylindrical cut-out with a diameter of \SI{28.5}{\mm} (red). The hole was either left empty or filled with PMMA, bone or adipose tissue surrogates. }
\label{fig:ExpSetUp}
\end{figure}

All experiments were performed at the Heidelberg Ion Beam Therapy Center (HIT), Germany, with the experimental setups shown schematically in figure \ref{fig:ExpSetUp}.
A cylindrical PMMA phantom of approximately human head size ($\textrm{radius} = \SI{16}{\cm}$, $\textrm{height} = \SI{9}{\cm}$) was placed in the isocenter of the treatment room. A two-layered TimePix stack (cf. section \ref{sec:Timepix}) operated in time mode was mounted horizontally at distance $d$ from the center of the phantom to the surface of the first detector layer. Its vertical center was aligned with the center of the phantom, and the inter-layer distance was \SI{3.6}{\mm}. Measurements were taken at different angles $\theta$ of the center of the sensitive area from the beam axis. In order to facilitate the rotation of the detectors around the phantom, a rotation device that can be controlled remotely was designed. 

For the proof of principle, the new imaging modality was tested on a homogeneous PMMA phantom. The TimePix stack was placed at a distance $d$ of \SI{10.3}{\cm} from the center of the phantom, at the angles $\theta$ of \SIlist{-80;-60;-40;-30;-20;10;20;30;50}{\degree} from the beam axis. The cylindrical PMMA phantom was irradiated with a carbon ion beam with an energy of \SI{226}{\mega\electronvolt\per\atomicmassunit} and a focus of \SI{4.6}{\mm}. The beam intensity and the number of initial particles was adapted to the measurement angle to result in approximately $5 \cdot 10^5$ measured secondary particles. The reconstruction method is described in detail in section \ref{subsec:Algorithm}.

After the successful proof of principle, we studied the possibility to visualize phantom inhomogeneities. A PMMA cylinder of the same outer dimensions as the homogeneous phantom, but with a cylindrical cut-out with a diameter of \SI{28.5}{\mm} was used (see figure \ref{fig:ExpSetUp}). Four different measurement sets were acquired. First, the cut-out (shown in red in figure \ref{fig:ExpSetUp}) was filled with a PMMA insert to create a homogeneous phantom as reference ($\rho_{\text{PMMA}}=\SI{1.19}{\g\per\cm\cubed}$). In the second measurement it was left empty ($\rho_{\text{air}}=\SI{0.0012}{\g\per\cm\cubed}$), then filled with a Gammex SB3 bone surrogate ($\rho_{\text{bone}}=\SI{1.82}{\g\per\cm\cubed}$) and lastly filled with a Gammex AP6 adipose tissue surrogate ($\rho_{\text{adipose tissue}}=\SI{0.92}{\g\per\cm\cubed}$) \footnote{www.sunnuclear.com/solutions/diagnostic/subcat/ct\_solutions/ct\_electron\_density\_phantom}. Two TimePix stacks were formed to speed up the experiment by measuring at two angular positions simultaneously. We placed the first stack at angles \SIlist{-50;-30;-10}{\degree} at $ d = \SI{10.3}{\cm}$, and mounted the second stack at angles \SIlist{40;60;80}{\degree} at $d = \SI{10.6}{\cm}$. The beam energy of \SI{226}{\mega\electronvolt\per\atomicmassunit} and beam width of \SI{4.6}{\mm} were the same for all experiments, while the intensity and particle number were again adapted to compensate for the decreasing yield of secondary ions with increasing angle form the beam axis.

\subsection{Three-dimensional reconstruction algorithm}
\label{subsec:Algorithm}

Compared to typical reconstruction problems such as CT-imaging, the monitoring of ion beams using secondary particles as studied in this work poses additional challenges. First of all, no dedicated large-area detection systems for the secondary ions are available. With the Timepix detector (c.f.\ \ref{sec:Timepix}) only a few detector positions can be measured in a reasonable time frame, leading to a small angular coverage of the forward hemisphere. The production of fragments is highly forward-peaked. This, combined with the small size of the detector results in a limited number of events per detector position at larger angles away from the beam axis. In addition, the measured particle tracks can originate from the whole path of the beam in the patient and thus arrive at the detector under multiple directions. Therefore a re-binning to a parallel geometry or collimation is not feasible. 

In a first approach, we investigated a simple volumetric backprojection method. This, however, resulted in prominent streaking artefacts. Due to the limited amount of data, analytical algorithms are not ideal for our purpose. Instead, iterative reconstruction algorithms were found to be better suited. We developed a three-dimensional reconstruction tool based on maximum likelihood expectation maximization (MLEM). 

The following outline briefly describes the principle of a MLEM reconstruction, based on \citep{Buzug}. The reconstruction problem can be expressed as $\mathbf{p} = \mathbf{A} \cdot \mathbf{f}^*$, where $\mathbf{p}$ is a vector of the measured quantity, $\mathbf{A}$ is a matrix corresponding to the imaging system, and $\mathbf{f}^*$ the vector of pixel values of the image that is a priori unknown. The aim of the MLEM reconstruction is to find the most probable solution $\mathbf{f}$. The reconstruction thus corresponds to the maximization of the likelihood $L(\mathbf{f})$.  A global maximum of the likelihood is found if the derivative is zero and the respective Hessian matrix is negative semi-definite. From these conditions, the following iteration rule can be deduced: 

\begin{equation}
f_r^{(n+1)} = \frac{f_r^{(n)}}{\sum\limits_{i=1}^{M} a_{ir}} \sum_{i=1}^{M} \frac{p_i }{\sum\limits_{j=1}^{N} a_{ij} f_j^{(n)}}a_{ir},
\label{eq:MLEM}
\end{equation}

where $p_i$ are the $M$ measured values, $N$ is the number of voxels, $a_{ir}$ are the components of the system matrix, $f_i^{(n)}$ is the current image estimate and $f_i^{(n+1)}$ is the updated image, i.e.\ the new estimate \citep{Buzug}.


We aim for a spatial resolution of \SI{1}{\cubic\mm} in the reconstructed image (resulting in $2.2 \cdot 10^6$ voxels for the phantom used). In addition, the small pixel pitch of the TimePix detector allows for a very high resolution in the track direction. However, this increases the size of the reconstruction problem to a system matrix with approximately $n \cdot 10^{16}$  entries ($n$ detector positions $\times$ $256^4$ possible tracks $\times$ $2.2 \cdot 10^6$ voxels), which is in practice not computable within reasonable computation times. We therefore developed a modified reconstruction method. To reduce the size of the problem, we optimized the data organization. We list all tracks instead of all possible pixel combinations,  which reduces the size of the problem by a factor of $10^4 - 10^5$. Additionally, it allows the computation to be carried out subsequently for each measurement position and track, leading to the new iteration rule: 

\begin{equation}
	f^{(k+1)}_j=\frac{f^{(k)}_j}{\sum\limits_{i=1}^{m}{a_{i,j}}} 	\cdot \sum\limits_{\text{angles}} 
	\left\lbrace \sum\limits_{\text{tracks}}
	\left[ \sum\limits_{i\in \text{track}} \frac{1}{\sum\limits_{j'\in \text{track}} {a_{i,j'}}f^{(k)}_{j'}} 
	\cdot a_{i,j} \right] \right\rbrace.
\end{equation} 

We implemented the algorithm outlined above in a C++ programme. To further decrease the computation time, the normalization to the system geometry, $1/{\sum_{i=1}^{M} a_{ir}}$, is precalculated. Under the given iteration rule, the likelihood is monotonously increasing. However the images tend to become noisy since a noisy image corresponds best to the noisy measurement. Therefore, a termination criterion different form simple convergence is chosen, depending on the application.


\section{Results}

\subsection{Proof of principle}
\label{subsec:ResRecfirst}

\begin{figure} 
\includegraphics[width=\textwidth]{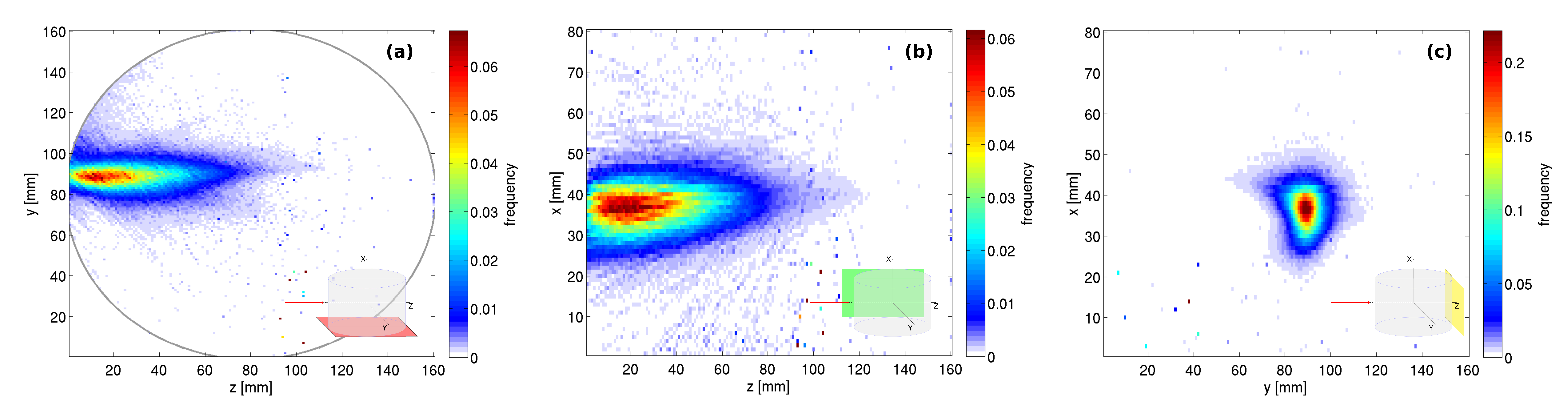}
\caption{Projections onto the three Cartesian orthogonal planes of the reconstructed image after 5 iterations. The incident carbon ion beam with a range of \SI{91.6}{\mm} in PMMA is clearly visible in all projections. }
\label{fig:ResfirstMLEMprojections}
\end{figure}

The main objectives of this work are to demonstrate the general feasibility of a three-dimensional reconstruction of a beam image in the phantom through the fragmentation points as well as a first analysis of inhomogeneity visualization. For the first application of the reconstruction algorithm to experimental data, the homogeneous phantom was irradiated with a carbon ion pencil beam and the emerging tracks were registered at nine discrete angular positions between \SI{-80}{\degree} and \SI{50}{\degree} (cf.\ sec. \ref{subsec:Experiments}). The three-dimensional image of the beam in the phantom was reconstructed as described in section \ref{subsec:Algorithm}. The projections of the image volume onto the yz-, xz- and yz-planes are shown in figure \ref{fig:ResfirstMLEMprojections}. Figure \ref{fig:ResfirstMLEMprojections} depicts the reconstructed image volume after 5 iterations. This number of iterations was chosen based on visual judgement as the best compromise between limited streaking artefacts and reduced noise. 

In all three images, the course of the primary carbon ion beam in the phantom is clearly visible. The signal is most prominent on the central axis of the phantom up to the phantom center. This corresponds to the expected beam region, since the ions were directed at the phantom center and stopped in \SI{91.3}{\mm} of PMMA. Only few traversing tracks behind the beam area can be seen. The limitation of the signal to the phantom volume is an intrinsic property of the algorithm, since the initial image estimate is set to zero outside of the object. The arising noise is on the one hand due to the iterative algorithm itself. In addition, the normalization to the detection geometry results in an increase of the noise. The detectors only cover a small part of the forward hemisphere, and the normalization image is a back projection of all possible tracks. Thus, the fields of view of the individual detector positions overlap partially, and interferences and border effects can occur. Especially patterns such as the prominent points in the lower center of the yz-projection (figure \ref{fig:ResfirstMLEMprojections}a) are artefacts of the normalization image. 
	
The shape of the beam area is not perfectly symmetric (c.f.\ fig.\ \ref{fig:ResfirstMLEMprojections}). This can be explained by the choice of the detector positions. The detector was placed, amongst others, at small angles from the beam axis with overlapping fields of view, \SI{10}{\degree} and \SI{20}{\degree}. This results in the slightly curved form of the beam area and the signal tail in the center of the phantom. 

It can also be observed that the reconstructed beam area does not lie exactly on the central axis of the phantom. This is likely caused by detector setup imperfections. Already small misalignments between the detector layers in the order of 3 pixels or a small rotation of the detector stack by \SI{3}{\degree} can noticeably distort the reconstructed image. 

It stands out that more signal in peripheral areas can be seen in the vertical projection (see figure \ref{fig:ResfirstMLEMprojections}b) than in the horizontal projection (see figure \ref{fig:ResfirstMLEMprojections}a). This can be traced back to the setup geometry. The detectors were placed on a horizontal ring around the cylindrical phantom. Therefore, the precision is better in the horizontal direction, also resulting in lower noise.

\begin{figure} 
\includegraphics[width=\textwidth]{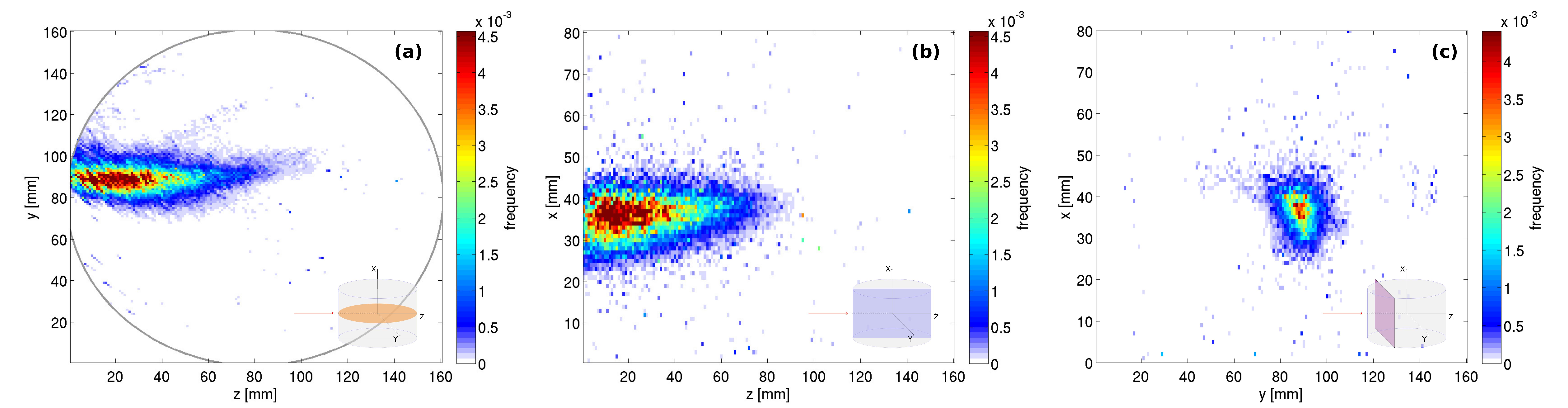}
\caption{Slices along the three Cartesian orthogonal planes of the reconstructed image after 5 iterations. The incident carbon ion beam with a range of \SI{91.6}{\mm} in PMMA is clearly visible in all slices. }
\label{fig:ResfirstMLEMslices}
\end{figure}

The projections onto the three coordinate planes already contain much information about the incident carbon ion beam. However, since they depict the sum of the three-dimensional image along one axis, small structures may be lost in the analysis. To further evaluate the performance of the modified MLEM algorithm, the \SI{1}{\mm} thick yz-, xz- and xy-slices through the center of the beam are given in figure \ref{fig:ResfirstMLEMslices}. Even though they show only a small portion of the available data, the beam area is clearly visible. As expected, the images appear noisier than the respective projections, but all features of 3D image are preserved.

\subsection{Visualization of target inhomogeneities}
\label{subsec:ResRecinhom}

\begin{figure}
	\includegraphics[width=\textwidth]{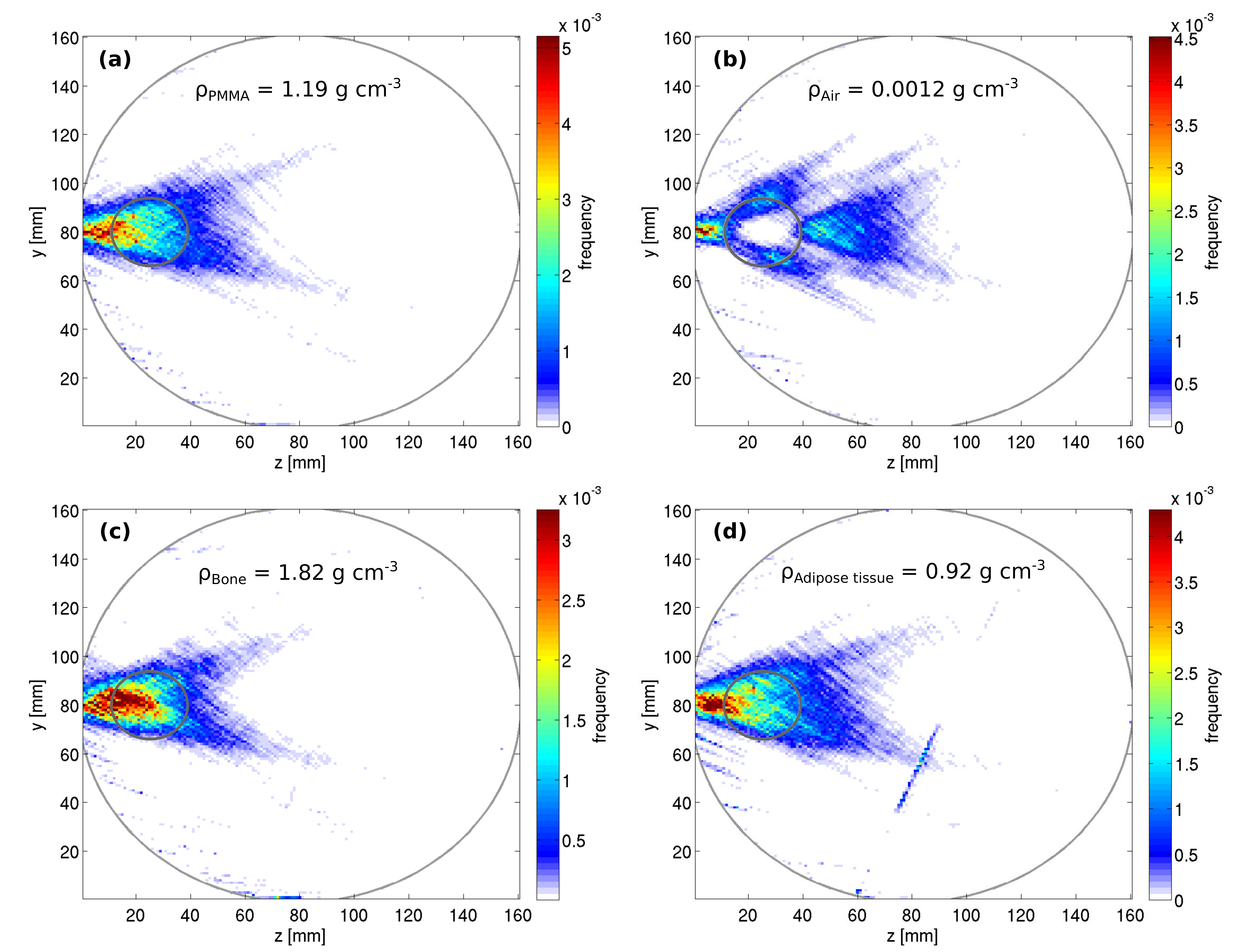}
	\caption{Horizontal, \SI{1}{\mm} thick slices through the phantom for the full PMMA phantom (a) and  the phantom with air (b), bone (c), and adipose tissue (d) inserts. The region of the inhomogeneity is marked by the small grey circle. Differences to the full phantom can be observed for all three cases.}
	\label{fig:ResMLEMInhomogeneities}
\end{figure}

In a next step, we applied the new data reconstruction method to a phantom with different inhomogeneities (c.f.\ figure \ref{fig:ExpSetUp}). We aim to investigate whether the method can directly visualize changes in the geometry. The resulting yz-slices through the phantom along the beam axis for all four inhomogeneities are presented in figure \ref{fig:ResMLEMInhomogeneities}. On a first glance, the air gap is by far the most obvious. Furthermore, an increase in signal in the region of the inhomogeneity with respect to the reference image  \ref{fig:ResMLEMInhomogeneities}a can be clearly seen for the bone insert, while slightly less signal can be observed for the adipose tissue surrogate. These tendencies are in accordance with the density differences and thus the differences in the fragmentation probability between PMMA and the inhomogeneities.

\begin{figure}
	\centering
	\includegraphics[width=.6\textwidth]{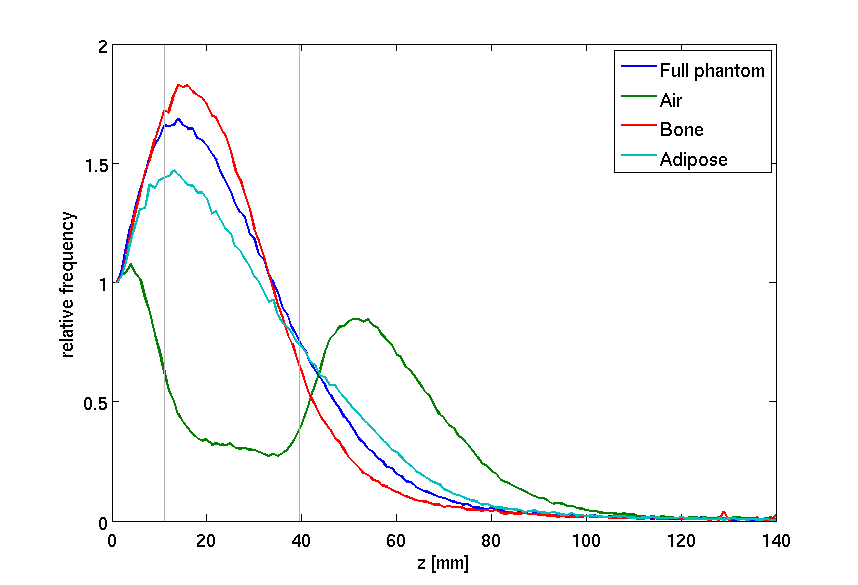}
    \caption{Projections of the reconstructed images onto the beam axis for different inhomogeneities. Changes in the area of the inhomogeneity (delimited by the grey lines) and the distal edge of the curves are clearly visible. The curves are normalized to the signal at a depth of \SI{0}{\mm}.}
    \label{fig:ResMLEMRangeProjInhom}
\end{figure}

Furthermore, we analysed the signal along the beam axis. To increase statistics, we summed over the central region in y-direction, from \SI{70}{\mm} to \SI{90}{\mm}. The curves are normalized to the entrance point, since the entrance material and beam energy are equal in all four cases and hence no differences in the entrance regions are expected. In the plot given in figure \ref{fig:ResMLEMRangeProjInhom}, the four phantom configurations are well distinguishable. 

The signal intensity within the inhomogeneity follows the material density, with the highest signal in the bone surrogate with \SI{1.82}{\g\per\cm\cubed}, followed by PMMA with \SI{1.19}{\g\per\cm\cubed}, adipose tissue surrogate  with \SI{0.92}{\g\per\cm\cubed}, and air with \SI{0.0012}{\g\per\cm\cubed}. The early decrease of signal in the bone insert is due to the shorter residual range of the primary particles. The non-zero signal in the air cavity is a result of smearing due to the algorithm, the cylindrical shape of the cavity, and scattering of the secondary particles in the phantom.


\section{Discussion}
\label{Discussion}

In this article, we presented a novel method for carbon ion beam radiotherapy monitoring based on single emerging secondary ion tracks. A three-dimensional reconstruction of the beam image from single secondary ion tracks was investigated. A homogeneous cylindrical PMMA phantom of human head-size was irradiated with carbon ion beams, and the secondary particles emerging from the phantom were tracked with a stack of parallel TimePix detectors. We developed a MLEM-based data reconstruction algorithm and showed its general applicability to this data. In addition, we demonstrated the possibility to visualize phantom inhomogeneities. 

Since this method only requires single tracks, the statistics is significantly higher than in interaction vertex reconstruction which requires two measured tracks from a single fragmentation event {\citep{Henriquet2012,Rescigno2014}. Compared to investigations of beam monitoring employing a monitoring detector \citep{Henriquet2012,Agodi2012,Piersanti2014}, the presented approach offers the advantage that no material is placed in the beam path in front of the patient. Scattering and fragmentation of the primary particles in this additional material would diminish the beam quality. In previous studies within our group \citep{Gwosch2013}, the 2D distribution of intersection points of the secondary particle trajectories with the beam plane was analysed for treatment verification. This introduces uncertainties due to the reduction of a three-dimensional problem to a two-dimensional plane, such as signal smearing due to the neglected finite beam width. Measurements at large angles from the beam axis as presented in \citep{Agodi2012,Piersanti2014} limit the influence of smearing due to the beam width in 2D approaches. However, this goes hand in hand with a loss of statistics by orders of magnitude. Our three-dimensional approach does not introduce smearing and allows the use of the forward peaked production of secondary ions. In contrast to beam delivery verification with PET, this new method offers the advantage that no extra time compared to the standard treatment work flow is required.

In this work, a three-dimensional beam image within the phantom was reconstructed from single measured secondary particle tracks. It mirrors the expected fragmentation point distribution, and provides information on both the beam extension as well as the phantom structure. In clinical applications of this monitoring technique, no extra dose to the patient would be required since the exploited secondary radiation is a by-product of the therapeutic irradiation.  Dedicated detectors for this method are not yet available, but the presented results are very encouraging for further developments and the application of trackers developed for high energy physics experiments. The TimePix detector proved to be a valuable tool in the presented approach due to its high detection efficiency for charged particles and high resolution combined with its small size, straightforward employment and the versatile application possibilities due to the different operation modes and stacking options. Even with the limited number of detectors available, the new modified MLEM algorithm shows promising results. Due to the significant influence of minor positioning uncertainties, a thorough positioning and calibration of the detection system is important to fully benefit from the new method.

We demonstrated the ability to detect inhomogeneities with the new data reconstruction algorithm. Materials with a large difference in density of $\Delta \rho \approx \SI{0.6}{\g\per\cm\cubed}$ can be visualized directly. Materials with a smaller density difference of $\Delta \rho \approx \SI{0.3}{\g\per\cm\cubed}$ could be clearly identified in the projection of the data onto the beam axis. In the presented experiments, the density changes were therapy-realistic, but the inhomogeneities were rather large in size, with a diameter of almost \SI{3}{\cm}. In future studies, the achievable spatial resolution of patient-realistic structure changes will be investigated. 

Previous studies for beam range monitoring \citep{Agodi2012,Piersanti2014,Gwosch2013} focussed on the distal edge of the distribution projected onto the beam plane. This shift in the distal edge due to the different carbon ion ranges in the inhomogeneities is also clearly observed in our 3D method (c.f.\ figure \ref{fig:ResMLEMRangeProjInhom}). The behaviour of the reconstructed beam image with changes in the beam energy will be investigated in future work, but the current results are promising for a use in carbon ion beam monitoring. 

The results presented above were achieved with measurements at only six defined angular positions of the detector around the phantom, irradiated with a single pencil beam. An increase in image quality and resolution is expected with a ring-like or helmet-like detection system covering a larger solid angle in a single measurement. The yield in measured particle tracks suggests that monitoring of scanned pencil beams could be possible with such a system. 

The very satisfactory results in the visualization of phantom inhomogeneities give rise to a number of possible applications of this novel method. With a larger angular coverage of the detectors and a larger radiation field, prominent structures within the patient could potentially be visualized. These landmarks could be registered to CT images and used as reference points for the beam delivery monitoring, thus leading to a combined patient imaging and beam delivery monitoring system.  Possibly, this could eliminate the need for a calibration of the range monitoring method in terms of the interaction point distribution and hence turn the approach into an absolute, real-time measurement technique. The presented results are a first step towards the feasibility of such an approach. Detailed studies with more patient-realistic setups are planned for the future.

\section{Conclusion}
\label{Conclusions}

We presented the successful reconstruction of a three-dimensional image of a carbon ion beam image in a phantom. We developed a modified MLEM algorithm to reconstruct the fragmentation point distribution from single secondary ion tracks measured behind the phantom. As a proof-of-principle, the reconstructed image in a homogeneous phantom corresponds well to the distribution expected from the beam energy and shape. In addition, density changes in the phantom down to $\Delta \rho \approx \SI{0.3}{\g\per\cm\cubed}$ could be visualised. Future work includes the investigation of more patient-realistic phantom set-ups and a thorough calibration of the positioning system. The method would benefit from larger area detectors, but already the results with the TimePix detector presented in this work are promising for carbon ion radiotherapy monitoring. 

\ack
We thank Tanja Gaa and Thorsten Heuser for the fruitful discussions, the DKFZ workshop for the design and construction of the dedicated rotational detector holder, and the HIT facility for the beamtime and special accelerator settings. MM is funded by the German Cancer Aid (Deutsche Krebshilfe). This work was performed in the frame of the Medipix collaboration.

\section*{References}  

\bibliography{AMR_Reco_References}

\end{document}